\documentclass[%
 reprint,
 amsmath,amssymb,
 aps,
]{revtex4-1}

\usepackage{graphicx}
\usepackage{dcolumn}
\usepackage{bm}
\usepackage{float}

\begin{document}

\title{Suppression of nonlinear Zeeman effect and heading error in earth-field-range alkali-vapor magnetometers}
\author{Guzhi Bao$^{1,2}$ }
\author{Arne Wickenbrock$^{1}$}
\author{Simon Rochester$^{3}$}
\author{Dmitry Budker$^{1,4,5,6}$}
\affiliation{$^{1}$Johannes Gutenberg-Universit{\"a}t  Mainz, 55128 Mainz, Germany\\
$^{2}$ Department of Physics, East China Normal University, Shanghai 200062, P. R. China. \\
$^{3}$ Rochester Scientific, LLC., El Cerrito, CA 94530, USA\\
$^{4}$ Helmholtz Institut Mainz, 55099 Mainz, Germany\\
$^{5}$ Department of Physics, University of California, Berkeley, CA 94720-7300, USA\\
$^{6}$ Nuclear Science Division, Lawrence Berkeley National Laboratory, Berkeley, CA 94720, USA}


\date{\today}

\begin{abstract}
The nonlinear Zeeman effect can induce splitting and asymmetries of magnetic-resonance lines in the geophysical magnetic field range. This is a major source of ``heading error'' for scalar atomic magnetometers. We demonstrate a method to suppress the nonlinear Zeeman effect and heading error based on spin locking.
In an all-optical synchronously pumped magnetometer with separate pump and probe beams, we apply a radio-frequency field which is in-phase with the precessing magnetization. In an earth-range field, a multi-component asymmetric magnetic-resonance line with $\sim 60\,$Hz width collapses into a single peak with a width of 22\,Hz, whose position is largely independent of the orientation of the sensor. The technique is expected to be broadly applicable in practical magnetometry, potentially boosting the sensitivity and accuracy of earth-surveying magnetometers by increasing the magnetic resonance amplitude, decreasing its width and removing the important and limiting heading-error systematic.


\begin{description}
\item[PACS numbers]
32.60.+i, 07.55.Ge, 32.30.Dx
\end{description}
\end{abstract}

\pacs{Valid PACS appear here}
\maketitle
\section{Introduction}\label{sec:Intro}
\begin{figure}[tbph]
\centering
\includegraphics[width=8.6cm]{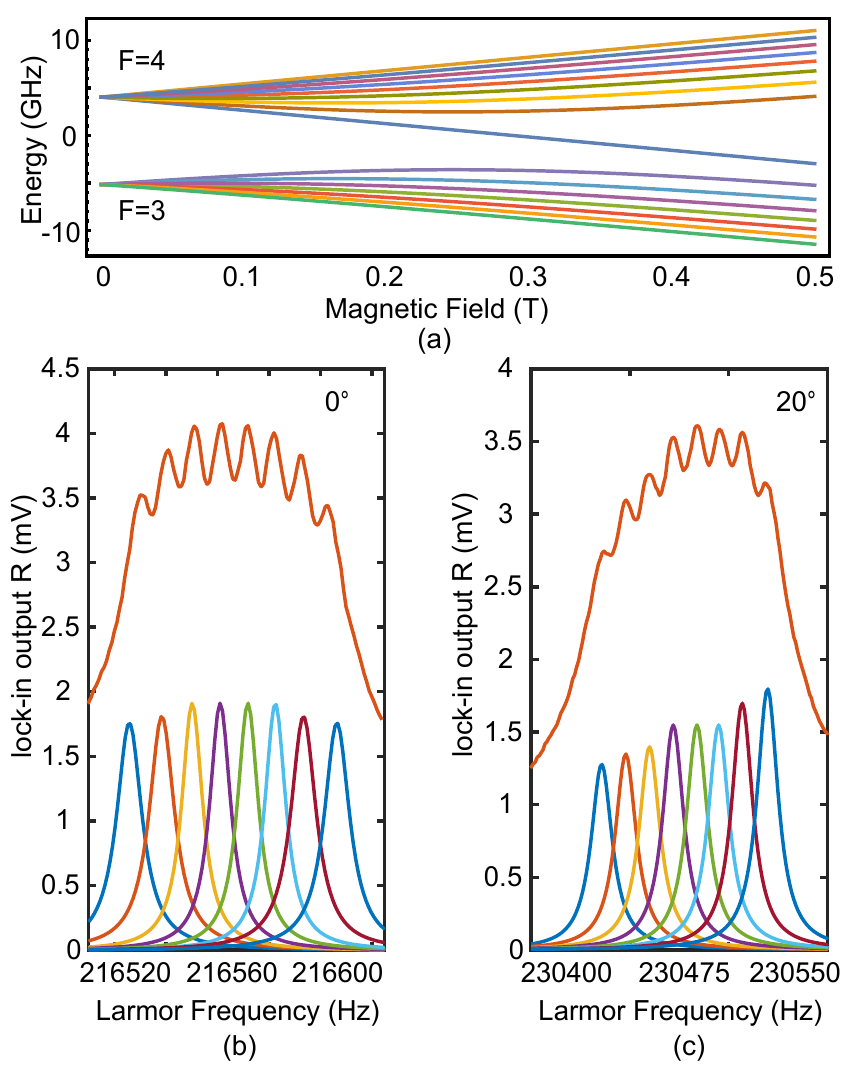}
\caption{(a) Hyperfine structure of the Cs ground state manifolds in an external magnetic field. In this work, we are concerned with fields that correspond to $\sim 10^{-4}$ of the shown range, where the NLZ effect is a small perturbation. (b) Optical-rotation signal amplitude (see Sec.\ \ref{sec:Experimental setup}) from the lock-in amplifier for a fixed pump modulation frequency of 216560\,Hz and a sweep of the magnetic field (given in Larmor frequency of the magnetic resonance neglecting NLZ shifts). The central field is $B\approx 62\,\mu$T. The data are fit with eight Lorentzian peaks arising due to the NLZ effect. (c) Same data collected with a $20^{\circ}$ tilt of the sensor and a pump modulation frequency of 230475\,Hz.}
\label{fig:hyperfineNLZ}
\end{figure}
High-sensitivity magnetometers are used in a wide variety of applications ranging from geophysics \cite{H.B.Dang(2010)} to fundamental physics \cite{G.Vasilakis(2009),I.Altarev(2009)} to medicine \cite{G.Bison(2009),C.N.Johnson(2013)}.  Alkali-metal-vapor atomic magnetometers have seen tremendous progress in recent years improving their sensitivities to below the $\textrm{fT}/\sqrt{\textrm{Hz}}$ level for submicrotesla fields \cite{H.B.Dang(2010),M.P.Ledbetter(2008),W.C.Griffith(2010),S.J.Smullin(2009),D.Budker(2007)}. However, in the geophysical field range (up to 100$\ \mu$T), the nonlinear Zeeman (NLZ) effect \cite{S.J.Seltzer(2007),V.Acosta(2006),K.Jensen(2009),li2016unshielded} can cause splitting of the different magnetic-resonance components and produce lineshape asymmetries. This leads to signal reduction and a spurious dependence of the scalar-sensor readings on the relative orientation of sensor and magnetic field. This important systematic effect is called heading error \cite{E.B.Alexandrov(2003),A.Ben-Kish(2010)} and becomes particularly troublesome in airborne and marine systems. 

Recently, NLZ shifts have been canceled using several different approaches: double-modulated synchronous optical pumping \cite{S.J.Seltzer(2007)}, high-order polarization moments \cite{V.Acosta(2006)}, and tensor light-shift effects \cite{K.Jensen(2009)}. Here, we introduce an alternative technique that is more generally applicable and easier to implement. It involves  ``locking" the atomic spins with an additional radio frequency (rf) field to suppress the NLZ effect and as a result also the heading error. Spin locking is often used in nuclear magnetic resonance (NMR) experiments to prevent precession or decay of the nuclear magnetization \cite{AlexanderJ.Vega(1985)}. In atomic systems, spin locking prevents splitting, shifts and lineshape asymmetries. In contrast to other schemes, spin-locked magnetometers are more robust against orientation changes.

\section{Theory}\label{sec:Theory}
\begin{figure}[tbph]
\centering
\includegraphics[width=8.6cm]{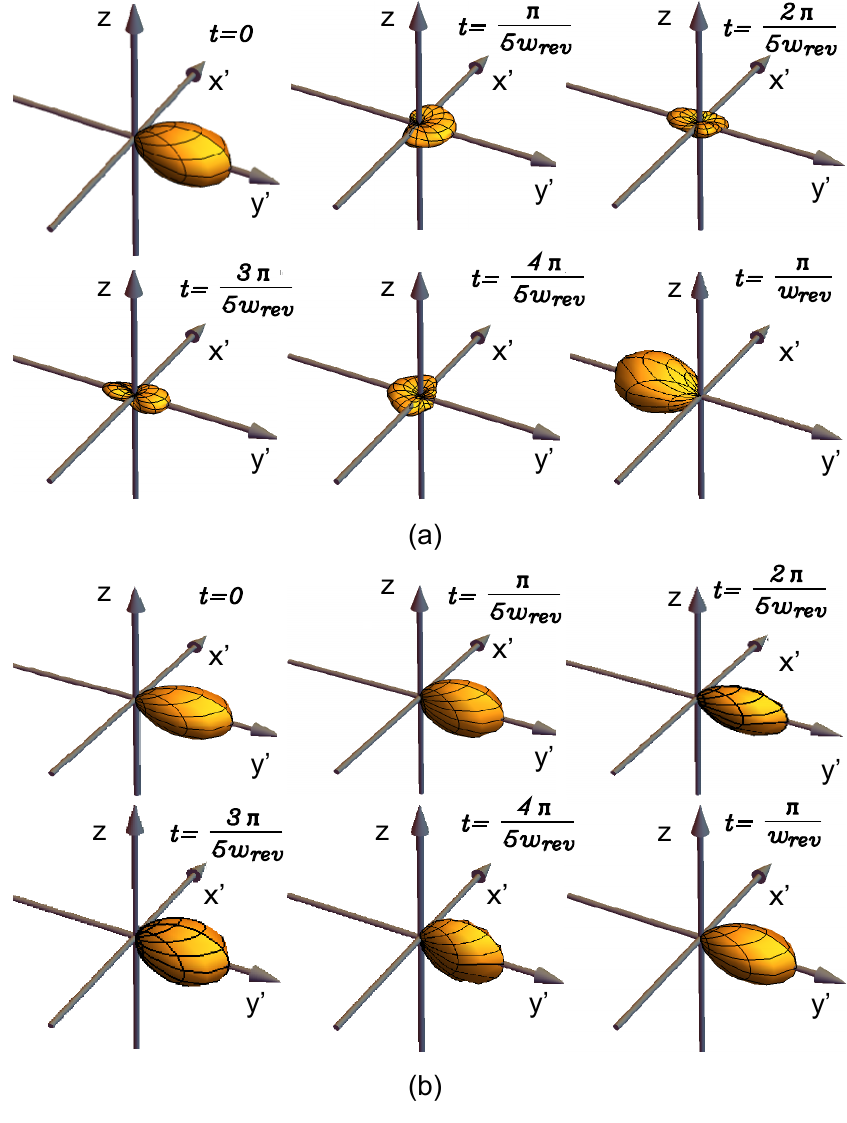}
\caption{Evolution of atomic polarization without (a) and with (b) spin locking in the rotating frame (spanned by $\hat{x'}$, $\hat{y'}$). The state is initially prepared stretched along the $\hat{y'}$-axis ($m_y=4$). A magnetic field is applied in $\hat{z}$-direction. The polarization state is represented with angular-momentum probability surfaces \cite{opticallypolarizedatoms} shown for particular phases of the Larmor precession of the spins.}
\label{fig:theory}
\end{figure}
For states with electronic angular momentum $J=1/2$, the energies of the magnetic sublevels $|m\rangle$  of a state with total angular momentum $F$ as a function of the magnetic field are given by the Breit-Rabi formula \cite{G.Breit(1931),opticallypolarizedatoms}:
\begin{equation}
E_{m}=-\frac{\Delta_{hf}}{2(2I+1)}-g_{I}\mu_{B}mB\pm\frac{\Delta_{hf}}{2}(1+\frac{4m\xi}{2I+1}+\xi^{2})^{1/2},
\end{equation}
where $\xi=(g_{J}+g_{I})\mu_{B}B/\Delta_{hf}$, $g_{J}$ and $g_{I}$ are the electronic and nuclear Land\'e factors, respectively, $B$ is the magnetic field strength, $\mu_{B}$ is the Bohr magneton, $\Delta_{hf}$ is the hyperfine-structure interval, $I$ is the nuclear spin, and the $\pm$ refers to the $F=I\pm1/2$ hyperfine components. The eigenenergies are shown in Fig.\,\ref{fig:hyperfineNLZ} (a) as a function of the magnetic field strength. It follows from Eq.\,(1) that the $\Delta m =\pm1$ Zeeman-transition frequencies in both manifolds are $m$-dependent with the difference dependent on $B$. In earth-field range, these contributions are already substantial. Therefore,  for the experiments presented here,  we expand the eigenenergies to second order in the magnetic field $B$. For the Cs\,$6^{2}S_{1/2}\,F=4$ state, the transition frequencies between adjacent Zeeman sublevels are then given by:
\begin{equation}
\omega_{F,m}\approx\frac{\mu_{B}B}{4\hbar}+\omega_{rev}(2m-1),
\end{equation}
where  $\omega_{rev} =(\mu_{B}B)^{2}/(16\hbar \Delta_{hf})$ is the quantum-beat revival frequency (see, for example, Ref. \cite{opticallypolarizedatoms}). In earth field (typically 50\,$\mu$T), this frequency for Cs is $\omega_{rev}=2\pi\cdot 3.3\,$Hz and comparable to the low-field magnetic-resonance width for the vapor cell used here. Therefore, our magnetometer is strongly affected by the nonlinear Zeeman effect at magnetic fields at the level of the earth's field. As shown in the data presented in Fig.\,\ref{fig:hyperfineNLZ} (b) and (c), the magnetic-resonance is split into eight peaks. The linewidth for each peak is 9\,Hz, but the effective linewidth of the NLZ-split magnetic resonance is 120\,Hz. The broadening of the resonance reduces magnetic sensitivity. Another effect, also visible in the data presented in Fig.\,\ref{fig:hyperfineNLZ} (c), occurs when the sensor is not properly oriented with respect to the magnetic field. Instead of a symmetric distribution of the peaks, a pronounced asymmetry appears when the magnetic field is not perpendicular to the pump and probe beams. This effect is called \textit{heading error} and leads to a systematic false reading of the magnetic field, which limits the usability of optical magnetometers, for example, in airborne-exploration applications at earth field.

If spins are initially oriented at an angle to the magnetic field, they precess around the magnetic field with frequency $\omega_{L}$. However, in the presence of NLZ, the evolution of the spins is more complex than just spin precession and is characterized by interconversion of different polarization moments leading to disappearance and periodic revival of orientation precessing at the Larmor frequency. The effect can be modeled as a periodic conversion among polarization moments, such as orientation-to-alignment conversion (OAC) \cite{opticallypolarizedatoms}, and is illustrated in  Fig.\,\ref{fig:theory} (a).  The main idea of our method is to provide a small ``locking'' magnetic field in the rotating frame along the direction of the main spin component. This prevents the spins from undergoing OAC, by forcing them to precess around this auxiliary field. If the Larmor frequency associated to this spin-locking field is much larger than the the revival frequency, $\omega_{rev}$, the nonlinear Zeeman effect is compensated.

Results of numerical calculations with the Atomic Density Matrix (ADM) package \cite{M.Auzinsh(2009),ADM} are shown in  Fig.\,\ref{fig:theory} without (a) and with (b) the spin-locking field to illustrate the respective evolution of the atomic polarization. In the absence of the locking field as shown in Fig. 2 (a), there are quantum beats corresponding to the conversion between various polarization moments leading to the collapse and revival of the spin orientation. When the locking field is applied as shown in Fig. 2 (b), atoms largely maintain their initial polarization while they undergo Larmor precession.
\section{Experimental setup}\label{sec:Experimental setup}
\begin{figure}[tbph]
\centering
\includegraphics[width=8.6cm]{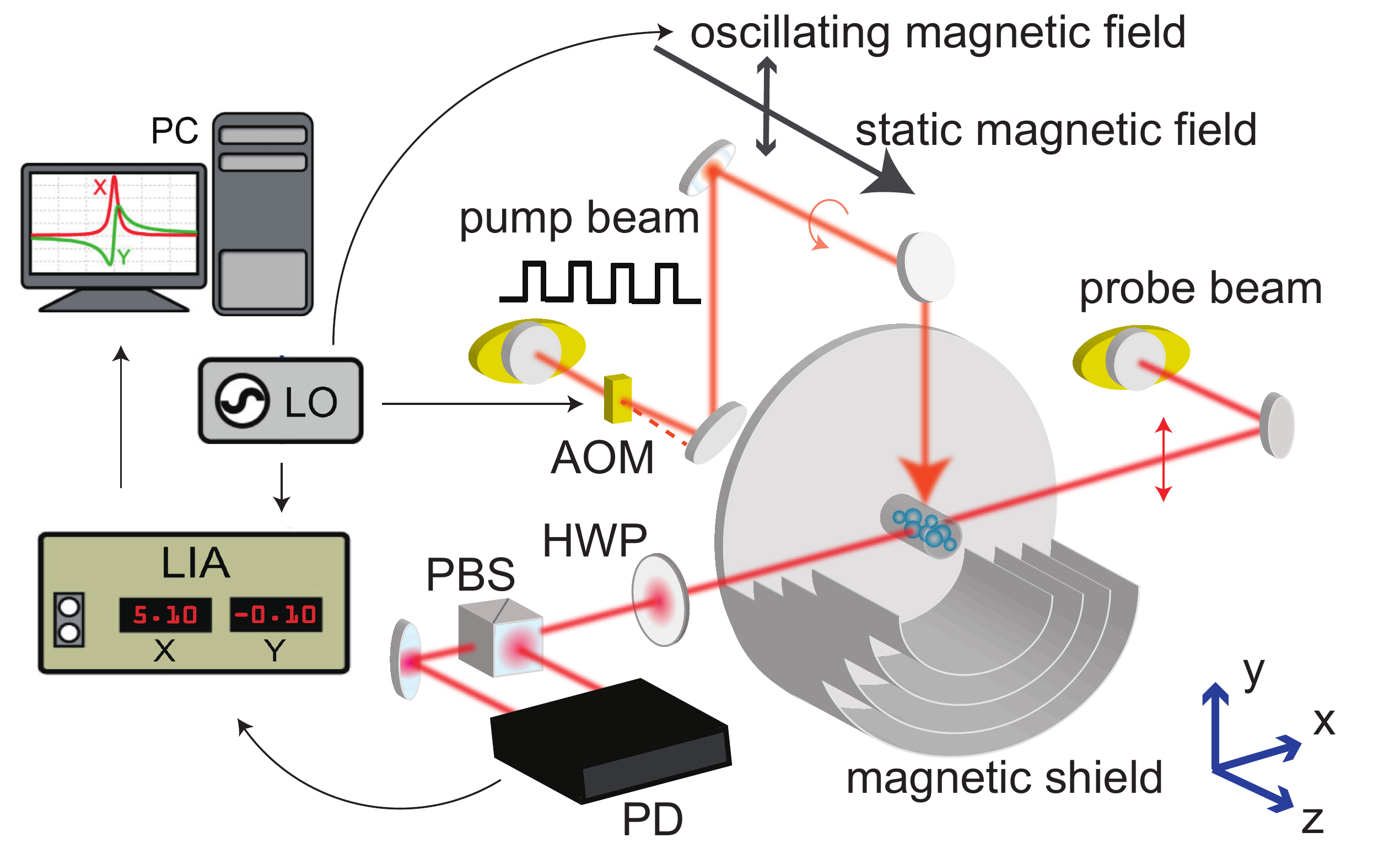}
\caption{Experimental setup. AOM: acousto-optic modulator used to pulse the pump beam; HWP: half-wave plate; PBS: polarizing beamsplitter; PD: balanced photodetector; LIA: lock-in amplifier; LO: local oscillator. Atoms are contained in vapor cell positioned in the center of the magnetic shield and are pumped and probed by laser beam under a static and oscillating magnetic field. Partial view of the magnetic shield is shown in the figure. }
\label{fig:apparatus}
\end{figure}
The experiment setup is shown in Fig.\,\ref{fig:apparatus}. A paraffin-coated cylindrical cell \cite{H.G.Robinson(1958)cell,M.A.Bouchiat(1964)cell,E.B.Alexandrov(1992)cell,E.B.Alexandrov(2002)cell} at room temperature with 4\,cm diameter and 5\,cm length containing $^{133}$Cs is enclosed within a four-layer mu-metal magnetic shield. A $-\hat{y}$-directed, circularly polarized pump beam is locked  to the Cs\,D2\,$6^{2}S_{1/2}\ F = 3\rightarrow\  6^{2}P_{3/2}\ F^{'}  = 4$ transition at 852\,nm with a dichroic atomic vapor laser lock (DAVLL) \cite{V.V.Yashchuk(2000)laserlock}. A static magnetic field up to 100$\,\mu$T is applied along $\hat{z}$; an oscillating magnetic field can be applied along $\hat{y}$ using coils within the inner shield. Additional DC magnetic fields can also be applied along $\hat{x}$ and $\hat{y}$ to tilt the total field away from $\hat{z}$ to study heading errors. The polarization of a $10\,\mu W$, $-\hat{x}$- directed, linearly polarized probe beam detuned by 4\,GHz to the blue from the Cs\,D2\,$F = 4\rightarrow  F^{'} = 5$ transition is measured with a balanced polarimeter upon transmission through the cell.

\begin{figure}[tbph]
\centering
\includegraphics[width=8.6cm]{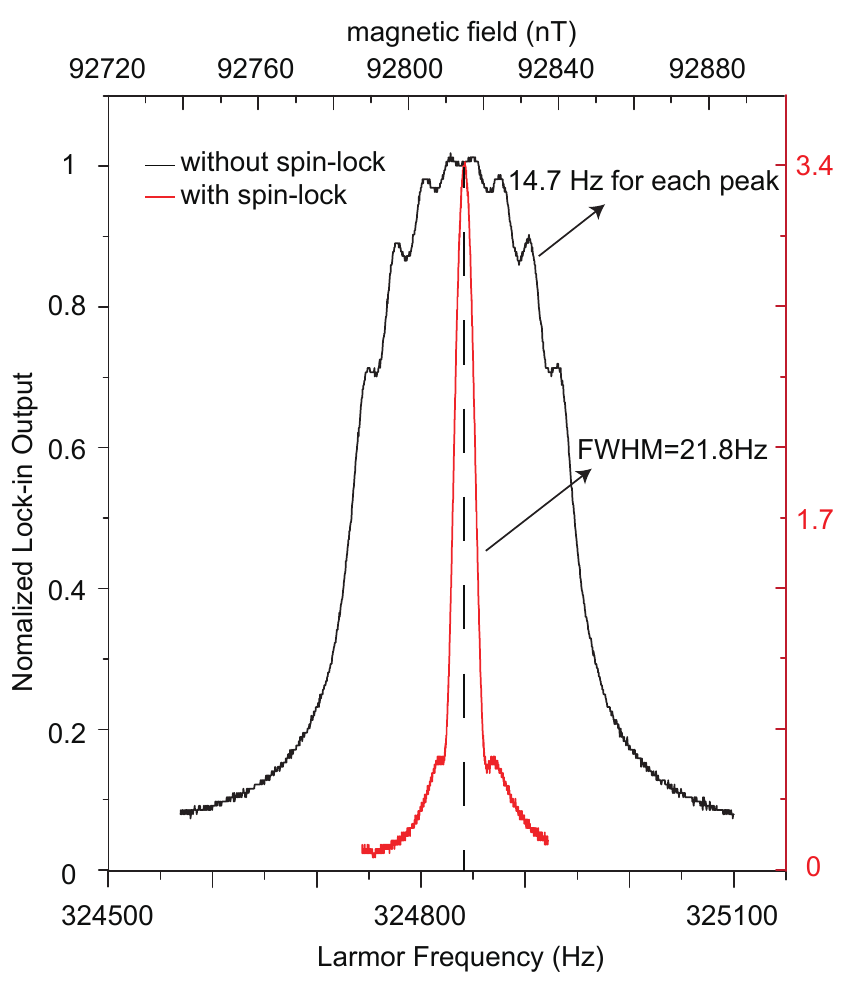}
\caption{Normalized magnetic resonance data for a pump-modulation frequency of 324840\,Hz as a function of the leading magnetic field along the $\hat{z}$ axis with (red, right vertical scale) and without (black, left vertical scale)  applied spin-locking rf field.}
\label{fig:1G}
\end{figure}

The pump beam is amplitude modulated (3\% duty cycle) with an acousto-optic modulator (AOM). The light power during the ``on'' part of the cycle is 50$\,\mu$W. We fix the modulation frequency at a particular value and scan the leading magnetic field until the polarization oscillation frequency is resonant with the Larmor precession, as detected via maximum polarization rotation amplitude of the probe beam \cite{D.Budker(2002)}. 
The signal from the balanced polarimeter is fed into a lock-in amplifier and demodulated at the modulation frequency. Examples of experimental scans are shown in Fig.\,\ref{fig:1G}. In the absence of the spin-locking rf field, the magnetic resonance is composed of eight Lorentzian peaks (each peak is 14.7\,Hz wide). Applying the rf field compresses all the different Lorentzians into a 21.8\,Hz wide central peak with a 3.4$\times$ increased amplitude.

To provide the spin-locking field, a sinusoidal current derived from the same function generator that modulates the pump-beam intensity is applied to the magnetic field coils in the $\hat{y}$-direction. 

\begin{figure*}[tbph]
\centering
\includegraphics[width=17.2cm]{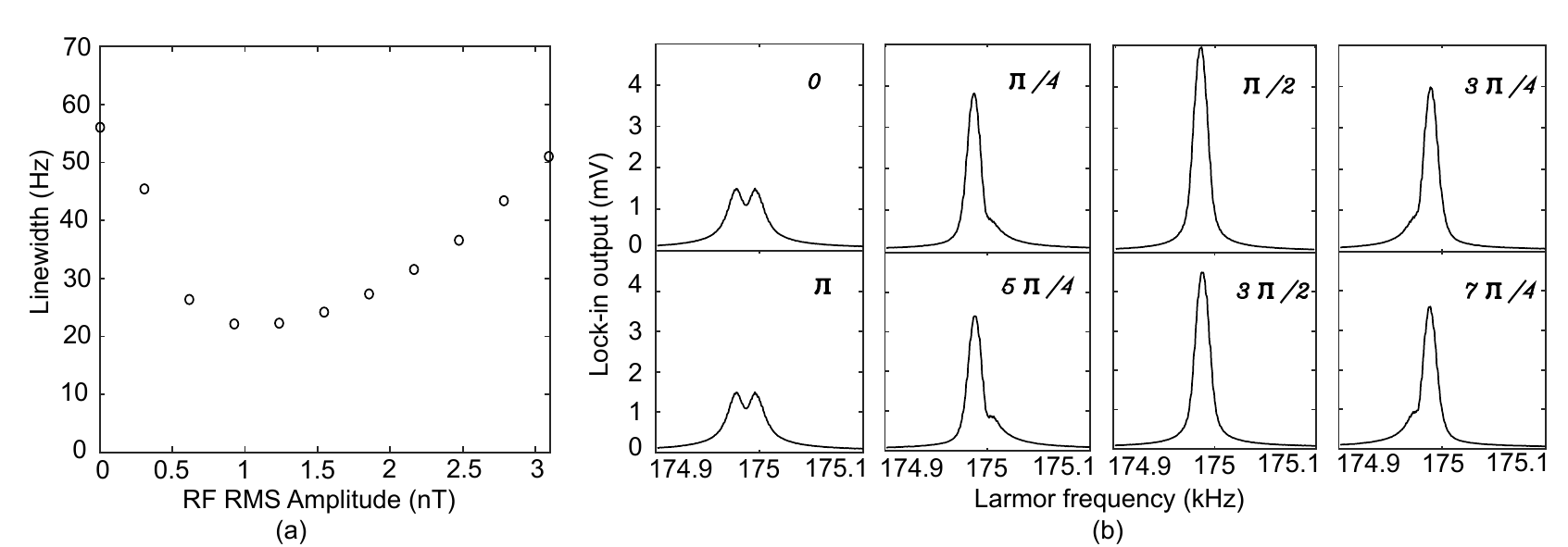}
\caption{(a) Linewidth change as a function of the applied spin-locking rf field amplitude. The minimum linewidth is 21.85\,(7)\,Hz. (b) Lineshape change for different phases $\phi$ between spin-locking rf and pump modulation. In the rotating frame a change in phase corresponds to a different angle between precessing magnetization and spin-locking magnetic field. Geometrically, a phase of $\pi/2$ corresponds to the field pointing along the magnetization. The respective phase $\phi$  is given in the top-right corner of each plot.}
\label{fig:compesationNLZ}
\end{figure*}
The oscillating magnetic field can be decomposed into the sum of two components rotating in the $x-y$ plane. One is co-rotating with the precessing spins, and the other (whose effect is non-resonant and will be neglected here) is counter rotating. In the rotating frame, the co-rotating field appears static. By using the internal control of local oscillator and adjusting the phase $\phi$ between the pumping pulses and the oscillating magnetic field, the effective direction of the static magnetic field in the rotating frame (spanned by $\hat{x'}$, $\hat{y'}$) can be changed like $\textrm{cos}(\phi)\hat{x'}+\textrm{sin}(\phi)\hat{y'}$, where $\hat{y'}$ points along the  magnetization. 

\section{Results}\label{sec:Results}

\subsection{\label{sec:level2}Suppression of the nonlinear Zeeman effect}
Figure\,\ref{fig:1G} shows the amplitude of the lock-in output as a function of the leading magnetic field around 92\,$\mu$T with the pump-laser modulation frequency fixed at 324840\,Hz. The data are presented without and with applying the spin-locking field in black and red, respectively. For this field, the spin-locking field is applied along $\hat{y}$ and has a 5.4\,nT RMS amplitude. 
The phase of the spin-locking field with respect to the pump modulation signal is optimized by maximizing the lock-in signal (R) in the center of the resonance. With spin-lock, the amplitude of the optical rotation signal is 3.4 times larger, while the effective linewidth of the central peak is an order of magnitude smaller. If the atomic spin is locked, ideally only one Larmor frequency exists in the system. The magnetic resonance should have the same amplitude and linewidth at low and high fields. Experimentally, however, the linewidth is broader [21.85\,(7)\,Hz] at high fields than at low fields [3.94\,(4)\,Hz]. We attribute this difference to power-broadening by the oscillating field and uncompensated magnetic-field gradients present when high fields are applied. 

Figure\,\ref{fig:compesationNLZ}\,(a) shows the effective linewidth of the magnetic resonance at earth field (50\,$\mu$T) as a function of the applied spin-locking field amplitude. We observe a minimum of the linewidth for an rf amplitude of 1\,nT where the precession frequency in the locking field is comparable to the revival frequency. Figure\,\ref{fig:compesationNLZ}\,(b) demonstrates the magnetic resonance lineshape for different directions of the spin-locking field as varied via the phase $\phi$. As expected, the best results are achieved for $\phi=\pi/2$ i.e. when the locking field is along the direction of the precessing spins. These results are confirmed by simulation based on the ADM package.
\begin{figure*}[tbph]
\centering
\includegraphics[width=17.2cm]{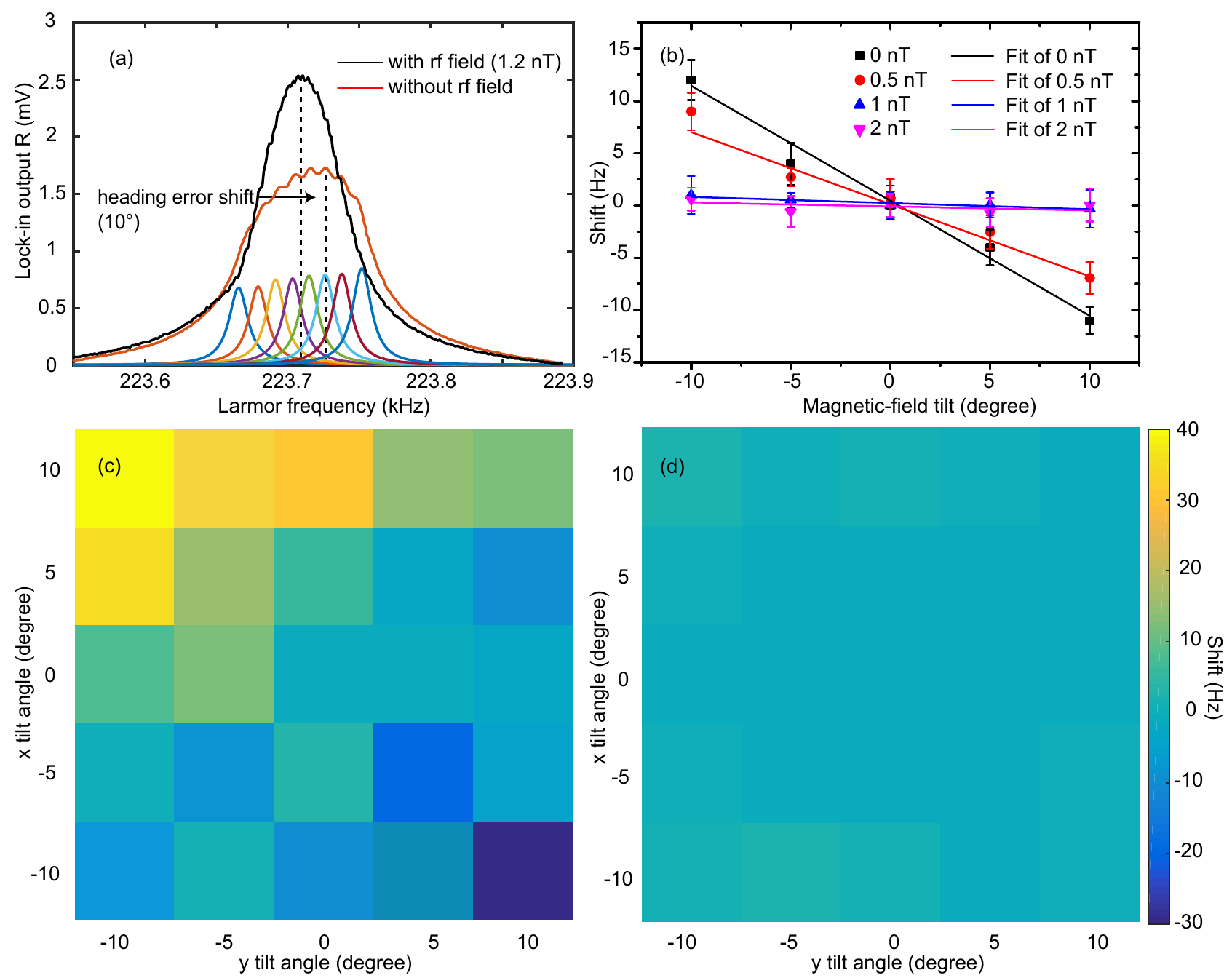}
\caption{
(a) Optical-rotation signal amplitude for a fixed pump modulation frequency of 223700\,Hz while scanning the magnetic field (expressed in Larmor frequency). Red curve shows the effect of heading error for a $10^{\circ}$ misalignment of the magnetic field axis to the direction of the probe beam (the $\hat{x}$-axis). A tilt in the field seems to shift the center of the resonance to higher frequencies and causes a visible asymmetry in the lineshape. We fit the signal with eight Lorentz peaks. The center of the resonance is decided by averaging eight individual Larmor frequency. The black curve shows the magnetic resonance for the same misaligned magnetic field with an applied 1\,nT spin-locking field. The resonance is symmetric with the correct central frequency. 
(b) Observed shift of the magnetic resonance due to heading error as a function of the misalignment angle for different amplitudes of the spin-locking field. Without this, the heading error is about 1.1\,Hz/degree and it reduces for increasing spin-locking field amplitudes. (c, d) Contour map of the two-dimensional heading error shift without and with rf field, respectively.}
\label{fig:headingerrorcompesition}
\end{figure*}
\subsection{\label{sec:level2}Suppression of heading error}
\begin{figure}[tbph]
\centering
\includegraphics[width=8.6cm]{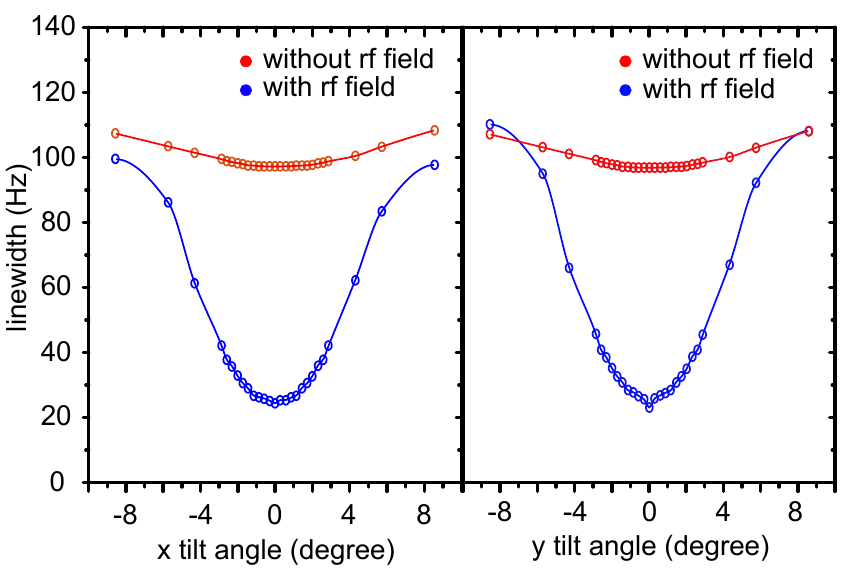}
\caption{(a) The linewidth change with tilt towards x direction without and with rf field. (b) The linewidth change with tilt towards y direction without and with rf field. Without rf field, the linewidth growth quadratically with tilt angle due to changing total field (tilt was done by adding a transverse component to the field). With rf field, the linewidth suppression diminishes with tilt angle.
}
\label{fig:changewithtiltangle}
\end{figure}
Figure\,\ref{fig:headingerrorcompesition}\,(a) shows the lock-in output for a tilt angle of 10$^{\circ}$ without and with spin-locking field. About 10\,$\mu$T misalignment field is applied along $\hat{x}$-direction tilting the overall magnetic field towards the probe beam. As shown in Fig.\,\ref{fig:hyperfineNLZ}, this shifts the weight of the individual magnetic resonances. This causes the combined lineshape to shift and to become asymmetric. The heading-error shift (difference between the central frequency and the maximum of the signal) is 12\,Hz. Applying an rf field, 
the single peak of optical-rotation signal with rf field appears at central frequency again. Figure\,\ref{fig:headingerrorcompesition} (b) shows the heading error as a function of the magnetic field tilt angle in the direction of the probe beam for different amplitudes of the rf field. Without the rf field, the heading error goes linearly with the tilt angle at a rate of 1.1\,Hz per degree.  The slope of the heading-error shift tends to zero with an increase in spin-locking rf amplitude.

To demonstrate the generality of the heading-error compensation, we vary the magnetic-field angle along both directions $\hat{x}$ and $\hat{y}$ in the range of $\pm\,10^{\circ}$. Figure\,6\,(c,d) show a contour map of the two-dimensional heading-error shift, without and with spin-locking rf, respectively. In the absence of rf field, the heading error ranges from -28\,Hz to 33\,Hz and with rf field, the shift is reduced over the full parameter space to be within $\pm\,0.7\,$Hz.

We notice that the linewidth is broadened by tilt toward both x and y. Figure\,\ref{fig:changewithtiltangle} shows the linewidth change with tilt angle. Within $\pm\,2$ degree, the spin locking is still working to make the linewidth narrower than 25 Hz. With larger tilt angle, the linewidth of signal is broadened and the amplitude decreases, but the heading error is still suppressed. 

 \section{\label{sec:level1}Conclusion and Outlook}
A method for suppressing the NLZ effect and heading error for magnetic fields in the range of the earth field using spin locking is demonstrated. An rf field along $\hat{y}$ direction is applied which effectively suppresses NLZ related broadening and heading error. The optimal spin-locking field corresponds to the Larmor frequency in the rotating frame comparable to the spin revival frequency; the phase is chosen such that the co-rotating part of the rf magnetic field is collinear with the precessing spins. We note that the sensitivity of earth-field magnetometers can be improved quadratically due to the increase in the signal amplitude and the reduction in the effective linewidth. The dependence of the linewidth on the tilt of the magnetic field and spin locking at a frequency different from that of pumping will be assessed in future work. The authors acknowledge support by the German Federal Ministry of Education and Research (BMBF) within the “Quantumtechnologien” program (FKZ 13N14439) and the DFG through the DIP program (FO 703/2-1). Guzhi Bao acknowledges support by the China Scholarship Council.
We thank Jason Stalnaker and Wenhao Li for helpful discussions.
\begin{equation}
\Delta E_{m}\approx \pm\frac{2}{2I+1}\mu_{B}mB\pm\frac{(\mu_{B}B)^{2}}{\Delta_{hf}}[1-\frac{4m^{2}}{(2I+1)^{2}}],
\end{equation}
\bibliography{ref.bib}
\end{document}